\documentclass{article}
\usepackage{spconf,amsmath,graphicx,hyperref,amsfonts}
\usepackage{xcolor}
\usepackage{cite}
\usepackage{booktabs}

\title{Direct Preference Optimization for Speech Autoregressive Diffusion Models}

\name{Zhijun Liu$^1$, Dongya Jia$^4$, Xiaoqiang Wang$^4$, Chenpeng Du$^4$, Shuai Wang$^{3,\dag}$, Zhuo Chen$^{4}$, Haizhou Li$^{2}$\thanks{$\dag$: Corresponding author}}
\address{
    $^1$School of Data Science, 
    The Chinese University of Hong Kong, Shenzhen \\
    $^2$School of Artificial Intelligence, 
    The Chinese University of Hong Kong, Shenzhen \\
    $^3$Nanjing University \quad
    $^4$ByteDance Seed \\
}

\begin{document}
\ninept
\maketitle

\begin{abstract}
Autoregressive diffusion models (ARDMs) have recently been applied to speech generation, achieving state-of-the-art (SOTA) performance in zero-shot text-to-speech. By autoregressively generating continuous speech tokens with next-token diffusion, these models offer a promising alternative to next-token prediction, avoiding the technical complexities associated with discrete speech tokenization. As a relatively new paradigm, research on reinforcement learning (RL)-based fine-tuning of speech ARDMs remains limited. In this paper, we propose Autoregressive Diffusion-Direct Preference Optimization (ARDM-DPO) to advance this research. By fine-tuning the recently proposed zero-shot text-to-speech model DiTAR with DPO, we achieve significant improvements in terms of speech expressiveness and robustness for long texts.
\end{abstract}


\section{Introduction}
\label{sec:intro}

A growing body of work in multimodal generation, including audio~\cite{ARDiT,DiTAR,SongBloom}, image~\cite{MAR,Orthus,NextStep1}, and video synthesis~\cite{CausVid,NOVA,MAGI1}, now employs autoregressive diffusion models (ARDMs) as the underlying architecture. ARDMs encode continuous modalities into sequences of continuous latent vectors (``continuous tokens”) and synthesize sequences by autoregressively predicting the next token with a diffusion model. Compared to next-token prediction over discrete token sequences~\cite{DiscreteTokens,DiscreTalk}, this next-token diffusion approach~\cite{LatentLM} preserves fine details while avoiding excessively long sequences, thanks to the compactness of continuous latent representations. Notably, recent studies~\cite{DiTAR,LatentLM,VibeVoice,DragonFM} applying ARDMs to speech generation report state-of-the-art zero-shot text-to-speech (TTS) performance.

Modern TTS systems trained on large datasets can sample from the data distribution with high fidelity. However, the generated speech may not always align with human preferences after pretraining. For example, when prompted with emotional speech, TTS models can still produce undesirable monotone outputs, requiring additional post-filtering by users~\cite{EmoDPO}. Preference alignment algorithms~\cite{InstructGPT,DPO,ImageReward} bias the output distribution of generative models toward human-preferred samples, making them an essential post-training step for powerful speech generation systems~\cite{SeedTTS,TTSDPO,INTP,CosyVoice3,DMOSpeech,RIO,SpeechAlign,UNO,DLPO,FPO,MPO}.

Direct Preference Optimization~\cite{DPO} (DPO) was originally proposed for aligning language models and was later adapted for fine-tuning diffusion models~\cite{DiffusionDPO}. In this work, we extend DPO to ARDMs~(ARDM-DPO) and apply it to fine-tune the recently proposed DiTAR model~\cite{DiTAR}, which achieves state-of-the-art performance in zero-shot TTS. DiTAR features a highly efficient ARDM architecture that separates computation for encoding the generation history from denoising the next token. To our knowledge, this is the first preference-alignment method tailored to ARDMs for TTS.

We evaluate ARDM-DPO on two benchmarks: (A) F0 variance for expressiveness and (B) text likelihood for robustness on hard texts challenging autoregressive TTS. ARDM-DPO nearly doubles F0 variance with minimal speaker-similarity loss and reduces CER by 25\%. Audio samples and further details can be found in the online supplement\footnote{\href{https://zjlww.github.io/ardm-dpo/}{https://zjlww.github.io/ardm-dpo/}}.

\vspace{-5mm}
\section{Background}
\label{sec:background}

\begin{figure*}[t]
    \vspace{-0.5cm}
    \centering
    \includegraphics[
        width=0.65\textwidth,
        trim=10 0 0 0, 
        clip,
    ]{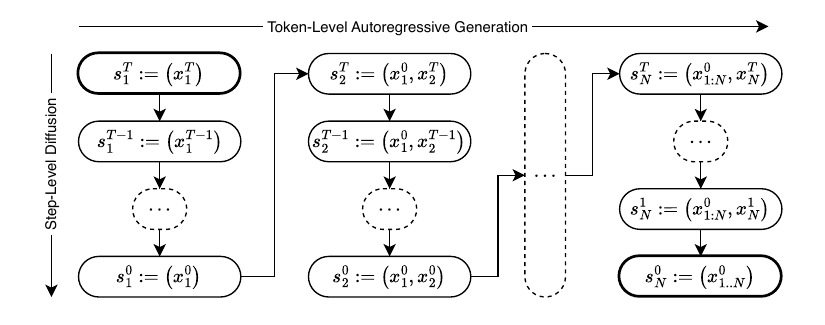}
    \vspace{-0.4cm}
    \caption{ARDM sampling viewed as a Markov chain. Each state contains both the history-generated tokens and the current noisy token. In this work, we define $x_{a:b} := \left \{ x_n: a \le n < b, n \in \mathbb Z \right\}$ and $x_{a..b} := \left \{ x_n: a \le n \le b, n \in \mathbb Z\right \}.$} 
    \vspace{-0.5cm}
    \label{fig:ardm_mdp}
\end{figure*}

\subsection{Diffusion Models}

Suppose $q(x_0)$ is the data distribution. For each diffusion time index $t \in \{1, \ldots, T\}$, define the positive decreasing sequence $(\alpha_t)_{t = 1}^T$ and the positive increasing sequence $(\sigma_t)_{t = 1}^T$. For each $t$ define the Gaussian perturbation distribution $q(x_t | x_0) := \mathcal N(x_t; \alpha_t x_0, \sigma_t^2 I_d)$. Define the noise perturbed distribution for each $t \in \{1, \ldots, T\}$ as
\begin{equation}
q(x_t) := \int q(x_0) q(x_t | x_0) \mathrm d x_0 .
\end{equation}
A well-trained diffusion model can be considered as a score estimator $\nabla_{x_t} \log q(x_t)$ trained with denoising score matching~\cite{ScoreSDE}. In the DDPM~\cite{DDPM,DDIM} sampler, the diffusion model generates samples starting from random noise $x_T \sim p(x_T) = \mathcal N(x_T; 0, I)$. Then it samples $x_{t - 1}$ given $x_t$ iteratively.
\begin{equation}
x_{t - 1} =  a_t x_t + b_t \nabla_{x_t} \log q(x_t) + c_t \epsilon,
\end{equation}
where $\epsilon \sim \mathcal N(0, I)$ is an independent noise, $(a_t)_{t = 1}^T$, $(b_t)_{t = 1}^T$ and $(c_t)_{t = 1}^T$ are positive sequences depending on $(\alpha_t)_{t = 1}^T$ and $(\sigma_t)_{t = 1}^T$.

\subsection{Autoregressive Diffusion Models}

For simplicity, we assume that the ARDM always generates $N$ continuous tokens. The ARDM sampling process can be viewed as a Markov chain, as illustrated in Fig.~\ref{fig:ardm_mdp}. Each state $s_{n}^t$ is indexed by the pair $(n, t) \in \{1, \ldots, N\} \times \{1, \ldots, T\}$, where $n$ is the token index and $t$ is the diffusion time index. Each state $s_{n}^t$ contains the history tokens already denoised $x^0_{<n}$ and the current noisy token $x_{n}^t$ that is being denoised, where we define $x_{<n}^0 := x_{1..n-1}^0$.

Given the state $s_n^t$, an ARDM estimates the conditional score $\nabla \log q(x_n^{t} | x^0_{<n})$ and transitions from $s_n^t$ to $s_{n}^{t - 1}$ when $t > 1$. Upon reaching state $s_{n}^0$, where a token is fully denoised, a random Gaussian noise $x_{n+1}^T$ is sampled, and a new DDPM sampling process starts from state $s_{n + 1}^T = (x_{\le n}^0, x_{n + 1}^T)$.

\section{ARDM-DPO}
\label{sec:method}

Let $\mathbf{x}$ denote an ARDM sampling trajectory that contains all intermediate states in Fig.~\ref{fig:ardm_mdp}. The reward function $r(x_{1..N}^0)$ is defined over the terminal states. For a trajectory $\mathbf{x}$, we define $r(\mathbf{x}) := r(x_{1..N}^0)$. Consider the following KL divergence-constrained policy optimization problem:
\begin{equation}
\max_{\pi} \left[ \mathbb{E}_{\pi(\mathbf{x})} \left[ r (\mathbf{x}) \right] - \beta D_{\mathrm{KL}}(\pi(\mathbf{x}) \| \mu(\mathbf{x})) \right],
\end{equation}
where $\beta > 0$ is the weight of the KL constraint. The KL-divergence constraint ensures that the learned policy $\pi(\mathbf{x})$ does not deviate too far from the reference distribution $\mu(\mathbf{x})$. The unique optimal policy $\pi_r$ given reward $r$ is~\cite{DPO,AWR}:
\begin{equation}
\pi_r(\mathbf{x}) := \frac{1}{Z_r} \mu(\mathbf{x}) \exp \left( \frac{1}{\beta} r(\mathbf{x}) \right),
\end{equation}
where $Z_r$ is defined as $Z_r := \int \mu(\mathbf{x}) \exp\left( \frac{1}{\beta} r(\mathbf{x}) \right) \mathrm{d} \mathbf{x}$. Then, we have:
\begin{equation}
r(\mathbf{x}) = \beta \log \frac{\pi_r(\mathbf{x})}{\mu(\mathbf{x})} + \beta \log Z_r.
\end{equation}
Marginalizing over all ARDM intermediate states gives:
\begin{equation}
\label{eq:r_as_log_ratio}
r(x_{1..N}^0) = \beta \mathbb{E}_{\pi_r(\mathbf{x} | x_{1..N}^0)} \left[ \log \frac{\pi_r(\mathbf{x})}{\mu(\mathbf{x})} \right] + \beta \log Z_r.  
\end{equation}
As in DPO, we assume that the log probability of human listeners preferring sample $x_{1..N}^0$ over $y_{1..N}^0$ is modeled by a Bradley-Terry model~\cite{DPO}:
\begin{equation}
\log P(x_{1..N}^0 \succ y_{1..N}^0) = \log \sigma \left( r(x_{1..N}^0) - r(y_{1..N}^0) \right).
\end{equation}
Given a dataset $\mathcal D$ of preference pairs, we can estimate the reward model parameters with maximum likelihood estimation:
\begin{equation}
\max_{r} \mathbb{E}_{x_{1..N}^0, y_{1..N}^0 \sim \mathcal{D}} \left[\log P(x_{1..N}^0 \succ y_{1..N}^0)\right].
\end{equation}
As a result of Eq.~\eqref{eq:r_as_log_ratio}, optimizing the reward model $r$ with maximum likelihood can be conducted without an explicit reward model by optimizing:
\begin{equation}
\mathbb{E}_{x_{1..N}^0, y_{1..N}^0 \sim \mathcal{D}} \left[ \mathcal{J}(x_{1..N}^0, y_{1..N}^0) \right],
\end{equation}
where
\begin{equation}
\begin{aligned}
&\mathcal{J}(x_{1..N}^0, y_{1..N}^0) := \\
&\log \sigma \Big( 
T N \beta \cdot \mathbb{E}_{\substack{\mathcal{U}(t), \mathcal{U}(n) \\ \pi(x_n^t, x_n^{t-1} | x_{\leq n}^0) \\ \pi(y_n^t, y_n^{t-1} | y_{\leq n}^0)}} \left[
\ell_{n}^t (\mathbf{x}) - \ell_{n}^t (\mathbf{y})
\right]
\Big),
\end{aligned}
\end{equation}
with $\ell_n^t (\mathbf{x})$ defined as:
\begin{equation}
\ell_n^t (\mathbf{x}) := \log \frac{\pi(x_{n}^{t - 1} | x_n^t, x_{<n}^0)}{\mu(x_n^{t - 1} | x_n^t, x_{<n}^0)}.
\end{equation}
As proposed in Diffusion-DPO~\cite{DiffusionDPO}, we assume that the distribution $\pi(x_n^t, x_n^{t - 1} | x_{\leq n}^0)$ can be approximated with $q(x_n^t | x_n^0) \cdot q(x_{n}^{t-1} | x_n^t, x_n^0)$. We also move the expectation over $\mathcal{U}(t)$, $q(x_n^t | x_n^0)$, and $q(y_n^t | y_n^0)$ from the inside of $\log \sigma$ to the outside with Jensen's inequality to obtain the approximate lower bound $\mathcal{L}$ for $\mathcal{J}$. The constant factor $T N$ is absorbed into $\beta$ in Eq.~\eqref{eq:elbo}. $\mathcal U(t), \mathcal U(n)$ are uniform distributions over all possible values.
\begin{equation}
\label{eq:elbo}
\begin{aligned}
&\mathcal{J}(x_{1..N}^0, y_{1..N}^0) \geq \mathcal{L}(x_{1..N}^0, y_{1..N}^0) := \\
& \mathbb{E}_{\substack{\mathcal{U}(t) \\ q(x_n^t | x_n^0) \\ q(y_n^t | y_n^0)}} \Big[
\log \sigma \Big( 
\beta \mathbb{E}_{\substack{\mathcal{U}(n) \\ q(x_{n}^{t-1} | x_n^t, x_{n}^0) \\ q(y_{n}^{t-1} | y_n^t, y_{n}^0)}} \left[
\ell_{n}^t (\mathbf{x}) - \ell_{n}^t (\mathbf{y})
\right]
\Big)
\Big].
\end{aligned}
\end{equation}
Notice that the expectation $\mathbb{E}_{q(x_n^{t - 1} | x_n^t, x_n^0)} [\ell_n^t (\mathbf{x})]$ can be written as the difference of KL divergences:
\begin{equation}
\label{eq:kl_divs}
\begin{aligned}
D_{\mathrm{KL}} \left( q(x_n^{t - 1} | x_n^t, x_n^0) \middle\| \mu(x_n^{t - 1} | x_n^t, x_{<n}^0) \right)\\
- D_{\mathrm{KL}} \left( q(x_n^{t - 1} | x_n^t, x_n^0) \middle\| \pi(x_n^{t - 1} | x_n^t, x_{<n}^0) \right).
\end{aligned}
\end{equation}
Without loss of generality, suppose that the ARDM is trained with the denoising objective. Then Eq.~\eqref{eq:kl_divs} is equivalent to:
\begin{equation}
\omega_t \left(-\|v_\theta(x_n^t, x_{<n}^0) - x_n^0\|_2^2 + \|v_{\mathrm{ref}}(x_n^t, x_{<n}^0) - x_n^0\|_2^2\right),
\end{equation}
where $\omega_t$ is a time-dependent weight. Finally, we arrive at the ARDM-DPO training objective:
\begin{equation}
\label{eq:obj}
\begin{aligned}
& \mathcal L(x_{1..N}^0, y_{1..N}^0) := \mathbb E_{\substack{\mathcal U(t) \\ q(x_n^t | x_n^0) \\ q(y_n^t | y_n^0)}} \Big [\log \sigma \Big (\beta\omega_t \mathbb E_{\mathcal U(n)} \\
\Big [ 
   &- \left \|v_\theta(x_n^t, x_{<n}^0) - x_n^0\right \|_2^2 + \left \|v_{\mathrm{ref}}(x_n^t, x_{<n}^0) - x_n^0\right \|_2^2\\
   &+ \left \|v_\theta(y_n^t, y_{<n}^0) - y_n^0\right \|_2^2 - \left \|v_{\mathrm{ref}}(y_n^t, y_{<n}^0) - y_n^0 \right \|_2^2
\Big ]\Big )\Big ].
\end{aligned}
\end{equation}
The DiTAR model used in our experiments is based on $v$-prediction, with continuous time $t \in [0, 1]$ where $t = 1$ corresponds to pure noise. We can derive the ARDM-DPO training objective for DiTAR as follows:
\begin{equation}
\label{eq:obj_vpred}
\begin{aligned}
& \mathcal L(x_{1..N_x}^0, y_{1..N_y}^0) := \mathbb E_{\substack{t \sim \mathcal U(0, 1) \\ x_{1..N_x}^1 \sim_{\mathrm{i.i.d.}} \mathcal N(0, I) \\ y_{1..N_y}^1 \sim_{\mathrm{i.i.d.}} \mathcal N(0, I)}} \Big [\log \sigma \Big (\\
& d^{-1}\beta \mathbb E_n\Big [ \left \|v_{\mathrm{ref}}(x_n^t, x_{<n}^0) - \dot x_n^t\right \|_2^2 - \left \|v_\theta(x_n^t, x_{<n}^0) - \dot x_n^t\right \|_2^2 \Big ]\\-
&d^{-1}\beta \mathbb E_n\Big [\left \|v_{\mathrm{ref}}(y_n^t, y_{<n}^0) - \dot y_n^t \right \|_2^2 - \left \|v_\theta(y_n^t, y_{<n}^0) - \dot y_n^t\right \|_2^2 
\Big ]\Big )\Big ].
\end{aligned}
\end{equation}
where $x_n^t = \alpha_t x_n^0 + \sigma_t x_n^1$ and $\dot x_n^t = \dot \alpha_t x_n^0 + \dot \sigma_t x_n^1$. $y_n^t$ and $\dot y_n^t$ are defined similarly. We dropped the time-dependent weight $\omega_t$ following the practice in Diffusion-DPO. In Eq.~\eqref{eq:obj_vpred} we compute $\mathbb E_n$ separately for $\mathbf x$ and $\mathbf y$, since the two trajectories can have different lengths $N_x$ and $N_y$. We normalized $\beta$ with $d^{-1} = 1/256$ in our experiments, with $d$ the dimensionality of each token in DiTAR.

\section{Experiments}
\label{sec:experiments}

\subsection{Common Setup}

\textbf{Base Model.} We fine-tuned a base DiTAR model with 0.4B parameters, pretrained on an internal corpus of around 280,000 hours of Chinese and English audio. The LM in the base model has 24 Transformer blocks, and the diffusion head contains 4 blocks. Each Transformer block has 1024 hidden dimensions and 16 attention heads.

\noindent
\textbf{Inference.} We used the same diffusion sampler for training sample generation and model evaluation. We use a 16-step DDPM sampler with a linear time schedule. We enabled LM Guidance~\cite{DiTAR}, which is similar to classifier-free guidance (CFG), with weight $w = 2$.

\noindent
\textbf{DPO Training.} All experiments were carried out on 32 A100 GPUs, with a local batch size of 1 pair and gradient accumulation of 32 steps. The effective batch size is 1024 pairs. We used the AdamW optimizer, with a fixed learning rate of $2 \times 10^{-6}$, weight decay $0.01$, $\beta_1 = 0.9, \beta_2 = 0.95$.

\noindent
\textbf{Objective Evaluations.} In all experiments, we report the word error rate (WER) using Whisper-large-v3 for English and the character error rate (CER) using Paraformer-zh for Chinese. Additionally, we calculate the cosine similarity of speaker embeddings (SIM) between the prompt and the generated audio using the WavLM-TDCNN model. All metrics were computed using Seed-TTS-Eval\footnote{\href{https://github.com/BytedanceSpeech/seed-tts-eval}{https://github.com/BytedanceSpeech/seed-tts-eval}}. We also report the token average KL divergence on the test set, which is defined as
\begin{equation}
d^{-1}\mathbb E_{\substack{\pi(\mathbf x), \mathcal U (n) \\ \mathcal U(t), q(x_n^t | x_n^0)}} \left \| v_\theta(x_n^t, x_{<n}^0) - v_{\mathrm{ref}}(x_n^t, x_{<n}^0)\right \|_2^2, 
\end{equation}
as a measure of the divergence between the fine-tuned model $\pi_\theta$ and the reference model $\mu$. For each objective metric, we report the average value across 8 random runs.

\noindent
\textbf{Subjective Evaluations.} Twenty listeners participated in pairwise listening tests, comparing audio from two TTS systems (e.g. A and B) on specific aspects (e.g., naturalness). For each pair, listeners chose ``A wins", ``B wins", or ``tie".

\subsection{Task A: Improving F0 Variance}

\noindent
\textbf{Task.} The fundamental frequency variance~(F0V) is strongly correlated with the perceived expressiveness of the generated speech. Optimizing F0V can effectively prevent the model from producing monotone responses.

\begin{figure}[htbp]
    \vspace{-0.3cm}
    \centering
    \includegraphics[
        width=0.36\textwidth,
        trim=0 0 -20 0, 
        clip,
    ]{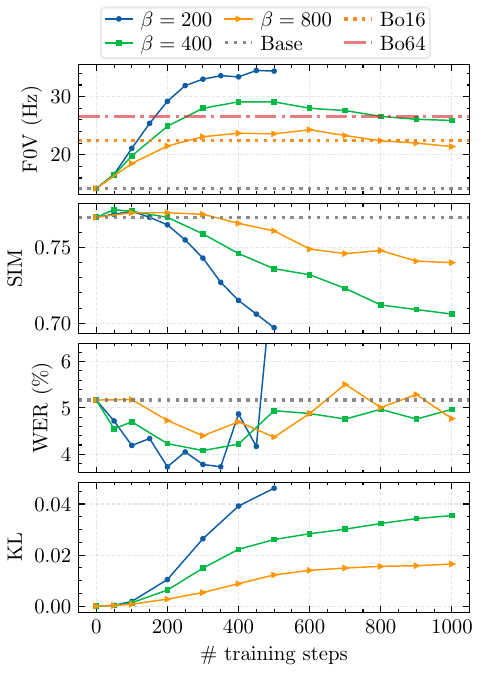}
    \vspace{-0.3cm}
    \caption{Trajectories of F0V, SIM, and WER of various models over 1000 training steps for Task A.}
    \vspace{-0.3cm}
    \label{fig:dpo_f0v}
\end{figure}

\noindent
\textbf{Preference Dataset.} We randomly sampled speech prompts and texts from the LibriTTS~\cite{LibriTTS} dataset. For each prompt-text pair, we generated 32 candidate responses using the base model. We then measured the F0V of these responses and selected the best and worst in terms of F0V to form a preference pair. In total, we collected 256k preference pairs, amounting to approximately 1,000 hours of speech.

\noindent
\textbf{Evaluation Dataset.} For evaluation, we randomly selected 38 prompt audios and target texts from different speakers in the LibriTTS \texttt{test-clean} subset.

\noindent
\textbf{Rejection Sampling Fine-Tuning.} We compare ARDM-DPO with rejection sampling fine-tuning (RAFT)~\cite{RAFT}. RAFT performs supervised fine-tuning (SFT) iteratively. In each RAFT iteration, we collect approximately 1,000 hours of speech continuations from the best policy found in the previous iteration. For each prompt, we sample 32 candidate responses and retain the one with the highest F0V. RAFT experiments use a batch size of 512 and a learning rate of $1 \times 10^{-5}$.

\begin{table}[h]
    \centering
    \vspace{-0.3cm}
    \begin{tabular}{ccccc}\toprule
         Method&  F0V $\uparrow$& SIM $\uparrow$ &WER $\downarrow$ &KL $\downarrow$\\\midrule
         Base Model& 
     14.2& 0.770&5.17 &---\\ 
 Best-of-16& 22.5& 0.770&4.74 &---\\
 Best-of-64& 26.6& 0.770&4.93 &---\\
 $\text{RAFT }^{\text{300 steps}}_{\text{iter }1}$& 18.3& 0.763&5.97 &0.057\\
 $\text{RAFT }^{\text{300 steps}}_{\text{iter }2}$& 19.7& 0.758&5.91 &0.230\\
 $\text{RAFT }^{\text{300 steps}}_{\text{iter }3}$& 20.1& 0.756&5.99 &0.237\\ 
 $\text{DPO}^{\,\text{200 steps}}_{\,\beta\text{ = 200}}$& 29.2& 0.765&3.73 &0.010\\ \bottomrule\end{tabular}
\vspace{-0.15cm}
    \caption{Selected objective evaluation results for Task A.}
    \label{tab:A}
\vspace{-0.3cm}
\end{table}

\noindent
\textbf{Grid Search for Optimal $\beta$.} In DPO training, $\beta$ controls the strength of KL regularization. We report results for $\beta \in \{200, \allowbreak 400, \allowbreak 800\}$. The evaluation results can be found in Tab.~\ref{tab:A} and Fig.~\ref{fig:dpo_f0v}. We observe that SIM gradually decreases throughout the training process for all values of $\beta$. This decrease is not caused by changes in prosody, as the best-of-K~(BoK) sampling results in Tab.~\ref{tab:A} indicate that an increase in F0V does not significantly reduce SIM. For larger values of $\beta$, the KL constraint is stronger, resulting in less degradation in SIM; however, the improvement in F0V is also smaller. We recommend applying early stopping to prevent significant quality degradation.

\begin{figure}[h]
    \centering
    \vspace{-0.3cm}
    \includegraphics[
        width=0.35\textwidth,
        trim=0 0 -20 0, 
        clip,
    ]{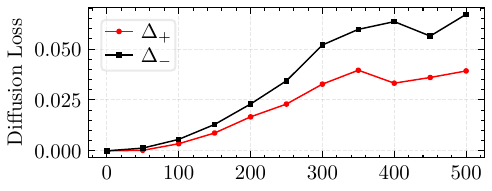}
    \vspace{-0.3cm}
    \caption{Change in diffusion loss of the winning and losing samples in Task A DPO with $\beta = 200$.}
    \label{fig:dpo_lossA}
     \vspace{-0.3cm}
\end{figure}

\noindent
\textbf{Diffusion Loss During DPO.} We visualize the changes in diffusion loss for the winning and losing samples ($\Delta_+, \Delta_-$) during training with $\beta = 200$ in Fig.~\ref{fig:dpo_lossA}. Although the training objective in Eq.~\eqref{eq:obj_vpred} is supposed to decrease the diffusion loss of the winning samples while increasing the diffusion loss of the losing samples, we observe that the model tends to increase both losses during training. A similar phenomenon has been observed in LLM DPO training~\cite{DPO3DP}. Investigating this behavior is left as future work.

\begin{figure}[h]
    \centering
    \vspace{-0.3cm}
    \includegraphics[
        width=0.5\textwidth,
        trim=10 0 -40 0, 
        clip,
    ]{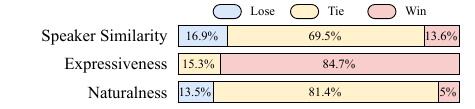}
    \vspace{-0.7cm}
    \caption{Results of the subjective evaluation: DPO vs. base model.}
    \label{fig:subjective_A}
     \vspace{-0.5cm}
\end{figure}

\noindent
\textbf{Subjective Evaluations.} For each of the 38 test cases, we generated three random responses from both the base model and the DPO model (200 steps, $\beta = 200$) for comparison. Evaluators assessed the response pairs based on three criteria: naturalness, speaker similarity to the prompt, and expressiveness. As shown in Fig.~\ref{fig:subjective_A}, DPO training slightly reduces naturalness and speaker similarity but significantly enhances perceived expressiveness.

\subsection{Task B: Improving Text Likelihood}

\noindent
\textbf{Task.} When evaluated on out-of-domain complex texts containing repetitions, autoregressive TTS models often make mistakes in audio-text alignment, such as missing or inserting words. Following prior work, we trained a phoneme-based CTC model~\cite{CTC} and used its negative log likelihood per phoneme (NLL) as a proxy for speech intelligibility. The CTC model consists of 6 transformer blocks, each with a hidden dimension of 1024 and 16 attention heads.

\noindent
\textbf{Preference Dataset.} The prompts were randomly sampled from DidiSpeech-2~\cite{DiDiSpeech}, a Chinese speech corpus consisting of 227 hours of recordings from 1,500 speakers. The texts were selected from a dataset of 100,000 long Chinese sentences, with randomly introduced repetitive phrases and clauses. For each prompt-text pair, we generated 16 candidate responses using the base model. We then calculated the CTC loss with our CTC model and selected the best and worst responses to form a preference pair. In total, we collected 430,000 preference pairs, totaling approximately 3,500 hours.

\noindent
\textbf{Evaluation Dataset.} We utilized the hard test set proposed in Seed-TTS~\cite{SeedTTS}, which contains 400 challenging test cases in Chinese with complex text. We excluded all speakers in the Seed-TTS-Eval hard set from the preference dataset, guaranteeing that they remained unseen during training.

\begin{table}[h]
    \centering
    { \fontsize{9}{11}\selectfont
    \begin{tabular}{ccccc}\toprule
         Method&  NLL $\downarrow$& SIM $\uparrow$ &CER $\downarrow$ &KL $\downarrow$\\\midrule
         Base Model& 
     0.55& 0.711&8.37 &---\\ 
 Best-of-8 (CER) & 0.39& 0.713& \colorbox{green!8}{4.99} &---\\
 Best-of-8 (NLL) & \colorbox{green!8}{0.27}& 0.712&6.79 &---\\ 
 $\text{DPO}^{\,\text{9000 steps}}_{\,\beta\text{ = 1600}}$& 0.32& 0.712&6.32 &0.009\\ \bottomrule\end{tabular}
 }
    \caption{Selected objective evaluation results for Task B.}
    \label{tab:B} 
\end{table}

\begin{figure}[htpb]
    \vspace{-0.3cm}
    \centering
    \includegraphics[
        width=0.36\textwidth,
        trim=0 0 -20 0, 
        clip,
    ]{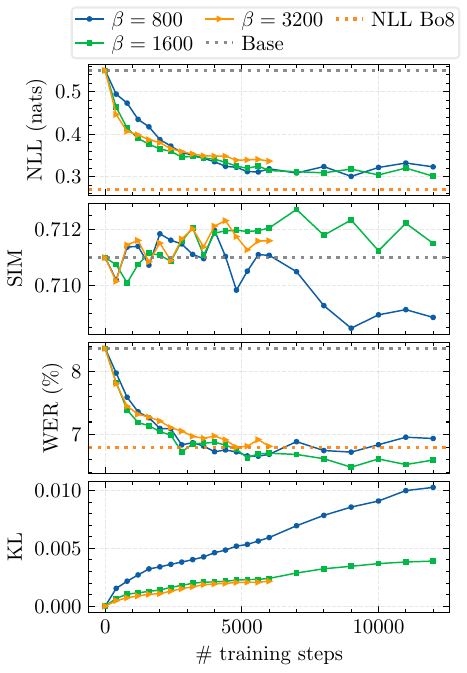}
    \vspace{-0.3cm}
    \caption{Trajectories of NLL, SIM, and CER of various models over 12,000 training steps for Task B.}
    \vspace{-0.3cm}
    \label{fig:dpo_nll}
\end{figure}

\noindent
\textbf{Grid Search for Optimal $\beta$.} We initially experimented with $\beta \in \{200, \allowbreak 400, \allowbreak \dots, \allowbreak 3200, \allowbreak 6400\}$ and observed that $\beta \leq 400$ led to increases in NLL and CER within 300 steps, while $\beta \geq 6400$ resulted in very slow optimization. Consequently, we focused on $\beta \in \{800, \allowbreak 1600, \allowbreak 3200\}$ for further analysis. The trajectories of NLL, SIM, and CER during training are shown in Fig.~\ref{fig:dpo_nll}. We found that the DPO model trained for 9000 steps with $\beta = 1600$ achieved the best performance, with a 25\% reduction in CER. Detailed results are provided in Tab.~\ref{tab:B}.

\noindent
\textbf{Subjective evaluations.} We randomly sampled 40 test cases from the test set and generated three random response pairs for each test case from the base model and the DPO model (9,000 steps, $\beta\text{ = 1600}$). Evaluators assessed all pairs for naturalness and speaker similarity. We find that the DPO model performs similarly to the base model. For naturalness, the lose/tie/win probabilities are 4.4\%, 88.7\%, and 6.9\%, respectively; for speaker similarity, they are 2.1\%, 94.3\%, and 3.6\%, respectively. This indicates good prior preservation of ARDM-DPO on Task B.

\section{Conclusions and Limitations}
\label{sec:conclusion}

In this work, we introduced ARDM-DPO, the first direct preference optimization method tailored for autoregressive diffusion TTS. Through comprehensive experiments on DiTAR, we demonstrate that ARDM-DPO achieves significant improvements in speech expressiveness and robustness on challenging long texts while maintaining speaker similarity and speech naturalness.

We observed that ARDM-DPO training on Task A is unstable and requires early stopping to avoid speech quality degradation. The underlying cause warrants further investigation. It is well known that the construction of the preference dataset in DPO plays a critical role in ensuring good performance~\cite{Tango2,TTSDPO,INTP}. We leave these directions for future work.

\bibliographystyle{IEEEbib}
\bibliography{main}

\end{document}